# Quantum Metrology Triangle Experiments: A Status Review

**Hansjörg Scherer and Benedetta Camarota**

Physikalisch-Technische Bundesanstalt Braunschweig, Bundesallee 100, 38116 Braunschweig, Germany

**Abstract**

Quantum Metrology Triangle experiments combine three quantum electrical effects (the Josephson effect, the quantum Hall effect and the single-electron transport effect) used in metrology. These experiments allow important fundamental consistency tests on the validity of commonly assumed relations between fundamental constants of nature and the quantum electrical effects. This paper reviews the history, results and the present status and perspectives of Quantum Metrology Triangle experiments. It also reflects on the possible implications of results for the knowledge on fundamental constants and the quantum electrical effects.

## 1. Introduction

The impending revision of the International System of units (SI) presents fundamental metrology with the most profound paradigm changes since the implementation of the SI by 11th General Conference on Weights and Measures in 1960 [1, 2]. The modern SI, based on the seven base units second, metre, kilogram, ampere, kelvin, mole, and candela, has been up to now very successful in ensuring worldwide consistency and uniformity of measurements. However, with scientific progress over the past half century, certain disadvantages are now apparent in the definition of the kilogram as the unit of mass in particular, but also in the definition of the electrical base unit ampere.

In the present SI, the kilogram is the last base unit still being based on a manufactured object, the international prototype of the kilogram, conserved and used by the International Bureau of Weights and Measures (BIPM) in France since 1889. Like any artefact, this platinum-iridium kilogram cylinder is susceptible to changes over time. Furthermore, the base electrical unit within the SI system, the ampere, is presently still defined in terms of mechanical units of mass, length and time via the laws of classical electromagnetism. This is unsatisfactory for two main reasons: firstly, the SI ampere is vulnerable to drift and instability from the kilogram artefact, and secondly, the electro-mechanical experiments needed to realise the SI electrical units are extremely difficult and require decades of effort. Moreover, under its present classical definition the ampere cannot be realised with an accuracy better than a few parts in $10^7$, which is not sufficient to meet the accuracy needs of routine electrical metrology, which requires 1 part $10^7$ now and will require even better in the future.

Since the 1980s, the Josephson effect and the quantum Hall effect, related to the fundamental constants $h$ and $e$ via the Josephson constant $K_J$ and the von Klitzing constant $R_K$, have proven their unexcelled precision and reproducibility of the order of 1 part in $10^9$ and better [3]. In order to exploit these effects for fundamental metrology, i.e. for the reproduction of the electrical SI units volt and ohm, and to benefit from the increased precision in electrical calibrations and measurements, in 1990 the 18th General Conference on Weights and Measures adopted the so-called conventional units for voltage and resistance ($V_{90}$ and $\Omega_{90}$), and defined fixed values for the Josephson and the von Klitzing constants ($K_{J-90}$ and $R_{K-90}$). Since then, the Josephson voltage standard (JVS) and the quantum Hall resistance (QHR) standard have been used for these metrological purposes with great precision, repeatability and ease [3, 4]. The conventional electrical units have achieved wide acceptance and are commonly used in science and industry. However, the definition of conventional units came with the price of a dilemma, since $V_{90}$ and $\Omega_{90}$ are not consistent with the SI definitions of the volt and the ohm.

Thus it is highly desirable to find a better, non-artefact-based definition of the kilogram, and a consequent definition of the ampere that could be realized in an easier and more precise way. This, together with the need to restore coherence to the SI system and enable practical unit realizations via direct traceability chains to invariants of nature, has driven efforts towards the re-definition of the SI units.

Thanks to scientific progress made in National Metrology Institutes (NMI) around the world during the last decades, the newly proposed SI unit definitions are entirely based on fundamental constants of nature and will consequently allow units realizations which are highly accurate and invariable over time [1, 2]. These definitions will be of explicit-constant type, i.e. the units will be defined by specifying exact values for certain fundamental constants. Of particular importance for electrical metrology are the new definitions of the kilogram, which will be connected to a fixed value of the Planck's constant $h$, and of the ampere, which will be based on a fixed value of the elementary charge $e$. As a natural consequence, these new definitions will remedy the dilemma of the conventional electrical units by making quantum standards suitable that are coherent with the SI. Consequently, the importance of the quantum electrical effects for the realisation and conservation of the units will be further strengthened.

A key point for the application of the Josephson and the quantum Hall effects for the future realization of the SI volt and ohm is the crucial assumption that the fundamental relations $K_J = 2e/h$ and $R_K = h/e^2$ are exact. Providing experimental support for this assumption is still an ongoing goal of modern fundamental metrology research, and its need has been repeatedly emphasised by the international Committee on Data for Science and Technology (CODATA) [5, 6]. Empirical information on possible corrections to the predicted fundamental relations can be provided by consistency tests, such as Quantum Metrology Triangle (QMT) experiments which involve the Josephson and the quantum Hall effect in combination with the single-electron transport effect as a third quantum electrical effect.

The purpose of this paper is to review the current status of QMT experiments, including developments since the publication of recent review papers on the topic [7, 8]. Special focus is laid on a particular realisation of the QMT, represented by the so-called 'Electron Counting Capacitance Standard' experiment.

## 2. Principle and implications of the QMT

In the mid-1980s rapid advances in electron-beam lithography techniques allowed the fabrication of sub-µm-sized metallic tunnel junction systems in which Coulomb blockade phenomena could be observed. This initiated the advent of Single-Electron Tunnelling (SET) experiments and brought ideas for corresponding metrological applications [9]. The first formulation of a QMT was presented by the Likharev group and published in 1985 in a paper on the theory of Bloch wave oscillations in small Josephson junctions [10].

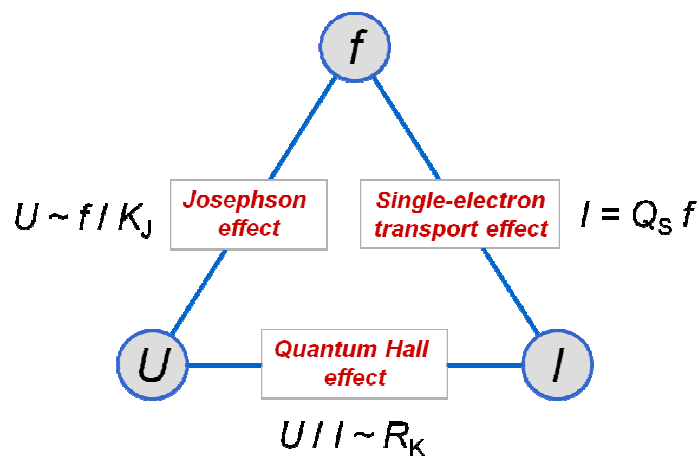

*Fig. 1: Original version of the QMT experiment similar to the first publication in 1985 [10], linking the three quantum representations of current, voltage and resistance.*

In this paper, for the first time the combination of three macroscopic quantum effects was proposed similar to Fig. 1) in order to investigate possible corrections to the underlying fundamental quantum relations.

About 15 years later, the original idea for the QMT experiment was newly formulated and interpreted on basis of the constituting relations for the three quantum electrical standards (Josephson voltage, quantum Hall resistance, and SET current) [11], given by

i)        $U_J = nf_J/K_J$ for the voltage produced by a Josephson voltage standard (JVS) operated at a frequency $f_J$ and on the nth voltage step,

ii)       $R_{QHR} = R_K/i$ for the resistance of a quantum Hall resistance (QHR) standard operated on the ith resistance plateau, and

iii)      $I_{SET} = Q_S f_{SET}$ for the current generated by an SET current standard device, driving charge quanta of value $Q_S$ at a frequency $f_{SET}$.

It is important to note that $K_J$, $R_K$ and $Q_S$ are introduced by these relations as 'phenomenological constants'. These are considered, indeed, *empirical quantities* whose values have to be determined experimentally by suitable electrical measurements. In particular, no relation of these constants to other fundamental constants of nature (like *e* and *h*) is assumed *a priori*.

Combining the three quantum effects by an experiment exploiting Ohm's law, i.e. by inserting i)-iii) into the relation $U = R I$, readily results in

$$K_J \, R_K \, Q_S = i \, n \, (f_J/f_{SET}). \qquad (1)$$

This relation represents the result of a QMT experiment. Such a result (as well as results from other, equivalent QMT variants, discussed later in this paper) tests the consistency of the quantum electrical effects by checking if the *product of the phenomenological constants* involved (the left side of equ. 1) is equal to a product of integer quantum numbers and a ratio of two frequencies (the right side of equ. 1). Here it is important to note that

- equ. 1) compares *dimensionless* products, i.e. all implications arising from QMT results are independent of the particular unit system chosen for the measured quantities, and

- the right side of equ. 1) is usually known with negligible uncertainty since frequencies (and their ratios, respectively) can be measured with very high accuracy by state of the art methods.

A number of standard theories for the quantum electrical effects exist relating $K_J$, $R_K$ and $Q_S$ to $e$ and $h$ [12, 13, 14, 15, 16]. These theories agree that these relations are given by

    ib)    $K_J = 2e/h$ for the Josephson constant,

    iib)    $R_K = h/e^2$ for the von Klitzing constant, and

    iiib)    $Q_S = e$ for the charge quanta constituting the electrical current in SET devices.

In fact some recent papers mention possible quantum-electrodynamical corrections to the von Klitzing and the Josephson constant in magnetic fields [17, 18], however the predicted dependencies are very weak, i.e. relative corrections of the order of $10^{-19}$ or less, so that they can be neglected under practical metrological aspects and with respect to the uncertainty levels that QMT experiments can reach.

In contrast to the famous relations ib) and iib), the relation $Q_S = e$ formulates a seemingly evident fact: namely, that the charge value carried by the charge quanta in solid-state devices is equal to the value of the electron charge in vacuum (i. e. the negative value of the elementary charge). The crucial question of whether many-body corrections to the electron charge exist in solid-state systems was first raised and treated in 1970 by Nordtvedt [19]. According to this work, the value of the electron-like charge quanta in solids is subjected to quantum-electrodynamic corrections, and the renormalized electron charge value in metals is higher than the vacuum value by a relative increase $\delta e/e$ of the order $10^{-10}$. Soon after that, however, several arguments were presented which cast doubt on the validity of Nordtvedt's conclusion [15, 16], stating that no such corrections apply. Presently this fundamental question is still considered an open topic [7, 8], and possible corrections cannot be ruled out *a priori*.

Regardless of the status of theoretical arguments, empirical tests like QMT experiments to verify the exactness of the relations ib) - iiib) at the highest possible confidence level are of uttermost importance for the application of quantum electrical effects in metrology and science.

To consider possible deviations from the ideal cases given by the relations ib) - iiib), corrections are commonly parameterized [11] according to

    ic)    $K_J = (1+\varepsilon_J)\, 2e/h$,

    iic)    $R_K = (1+\varepsilon_K)\, h/e^2$, and

iiic)    $Q_S = (1+\varepsilon_S)\,e$.

Inserting this into equ 1) leads then to the expression

$$K_J\, R_K\, Q_S\, /\, 2 \cong (1 + \varepsilon_J + \varepsilon_K + \varepsilon_S) \qquad (2)$$

in a first order approximation, i.e. assuming that the epsilon corrections each are much smaller than unity so that their products can be neglected.

Equ. 2) shows that if there are no corrections to any of the three involved quantum electrical effects (all epsilon corrections equal to zero), the QMT provides a consistency check by testing the relation $1 = 1$. Any result of a QMT experimental can be thus be expressed as

$$1 = 1 + \Delta_{QMT} \pm u_{QMT} \qquad , \qquad (3)$$

where $\Delta_{QMT}$ is the measured deviation from the expected $1 = 1$ relation, and $u_{QMT}$ is the relative standard uncertainty attributed to the result.

If $\Delta_{QMT} > u_{QMT}$, the experimental QMT result would imply that at least one of the three involved quantum effects has a correction term; however, this result would not allow to identify the effect. If $\Delta_{QMT} < u_{QMT}$, the experiment 'closes' the QMT, which means that evidence against corrections to the three involved quantum effects is provided on a confidence level of $u_{QMT}$. In this case though, the possibility of a cancellation between individual epsilon correction terms cannot be ruled out [11, 7].

## 3. Present knowledge of the values for the phenomenological constants

To assess the metrological impact boundaries of QMT experiments it is necessary to consider the present knowledge of the values of the phenomenological constants $K_J$, $R_K$, and $Q_S$, and their correction terms $\varepsilon_J$, $\varepsilon_K$ and $\varepsilon_S$. In the past, discussions on the QMT have formulated the ambitious ultimate target to reach a relative standard uncertainty $u_{QMT}$ of about one part in $10^8$ (see for instance [11]), or even state that this uncertainty level is necessary for significant metrological impact [20]. However, a careful and conservative analysis based on recent CODATA results [6] which follows the rationale published by Keller in 2008 [7] implies that the 'metrological impact threshold' for QMT experiments is significantly lower, namely at an uncertainty level of about few parts in $10^7$.

In 2010 CODATA performed the latest adjustment calculations of the fundamental constants, including an update of the adjustment calculations for possible corrections to the phenomenological constants $K_J$ and $R_K$ [6, 21]. The results were derived by least-squares adjustment calculations of the phenomenological constants based on input data from a wide variety of experiments, as described in earlier CODATA publications [22, 23, 5]. Some of these calculations were carried out *without* the assumption that the relations $K_J = 2e/h$ and $R_K = h/e^2$ are exact. These so-called 'relaxed conditions' were – according to the equations ic) and iic – considered by introducing adjustable correction factors $\varepsilon_J$ and $\varepsilon_K$ in the observational equations. The corresponding adjustment calculations then provided a set of 'best values' for these epsilon correction terms. According to the CODATA analysis from 2010 (see [6], pp 62, *Test of the Josephson and quantum Hall effect relations*) the values for the correction terms are (with all stated uncertainties here being "standard uncertainties" [6]):

- $\varepsilon_J = (15 \pm 49) \cdot 10^{-8}$, i.e. there is no significant correction to the predicted value of the Josephson constant at a confidence level corresponding to a relative uncertainty of about 5 parts in $10^7$

- $\varepsilon_K = (2.8 \pm 1.8) \cdot 10^{-8}$, i.e. there is a barely significant correction to the predicted value of the von Klitzing constant at a confidence level corresponding to a relative uncertainty of about 2 parts in $10^8$.

Interestingly, the correction factor for $K_J$ has a relatively high uncertainty of about 5 parts in $10^7$. This seems surprising since the Josephson effect nowadays is considered one of the best understood quantum electrical phenomena. The reason for this high uncertainty is due to a peculiarity that was already revealed in the CODATA report from 2006 [5]. Considering the fact that the value for $\varepsilon_J$ was mainly determined by different types of observational equations and experimental input data (see [7] for an extensive discussion), it was found that different 'routes' for the adjustment calculations led to strongly discrepant results for $\varepsilon_J$. Consequently, in order to obtain a result free of inconsistencies, additional adjustment calculations were performed with all sets of input data resulting in discrepant results deleted [23, 5, 6]. The 2006 adjustment then gave $\varepsilon_K = (2.4 \pm 1.8) \cdot 10^{-8}$ and $\varepsilon_J = (2.4 \pm 7.2) \cdot 10^{-7}$. Comparison with the new results of the CODATA analysis from 2010 [6] shows that the uncertainty for the correction factor to the Josephson constant now has slightly decreased from 7 to about 5 parts in $10^7$.

In 2008, a value for the third possible correction factor $\varepsilon_S$ was deduced by combining the results of a QMT experiment performed at NIST [24, 25] with results from Watt balance and calculable capacitor experiments (see [26] and references therein). This analysis gave

- $\varepsilon_S = (-9 \pm 92) \cdot 10^{-8}$, i.e. there is no significant correction to the predicted value of the charge quanta transported by SET devices at a confidence level corresponding to a relative uncertainty of about 9 parts in $10^7$.

The uncertainty for $\varepsilon_S$ here corresponds to the relative total uncertainty of the QMT experiment from NIST [25].

In summary: the relative uncertainty for a correction to $R_K$ is about 2 parts in $10^8$, for a correction to $K_J$ it is 5 parts in $10^7$, and for a correction to $Q_S$ it is 9 parts in $10^7$. Consequently, the implications of experimental QMT results are assessed as follows [7]: A QMT result with an uncertainty $u_{QMT}$ at about 1 part in $10^6$ (or higher) has to be interpreted primarily in terms of $\varepsilon_S$. An uncertainty in the range about 5 parts in $10^7$ and about 2 parts in $10^8$ would have impact on $\varepsilon_S$ and $\varepsilon_J$ together, keeping in mind that a QMT result cannot distinguish between them according to equ. 2). A result with $u_{QMT} < 2$ parts in $10^8$ would bear on the correction factors for all three quantum electrical effects.

This means that any QMT result with a relative total uncertainty at the level of about a few parts in $10^7$ can provide relevant input to future adjustments of the phenomenological constants. Such a result would contribute to reinforce with an empirical approach the theoretical models existing for the electrical quantum effects and their foundation as the basis for the future SI.

## 4. Implementation of various QMT experiments

At the time of the original formulation of the QMT (Fig 1), its experimental realization was not straightforward. In the early 1990s in fact, when SET devices started entering metrology applications, state of the art SET current sources were represented by single-electron pump or turnstile devices based on series arrays of metal-insulator-metal tunnel junctions [9, 27]. Due to inherent physical limits set by the statistics of the tunnelling process, the levels of quantized current achievable with these SET current sources could not exceed the range of about few picoampere. An SET current of 10 pA driven through a quantum Hall standard operating on the highest resistance plateau ($i = 1$, $R_{QHR} = R_K \cong 25.8$ k$\Omega$) results in a Hall voltage of about 40 nV. Measuring this voltage with a relative uncertainty $< 10^{-6}$ would require an accuracy of about $4 \cdot 10^{-14}$ V, exceeding the capabilities of the best JVS systems by orders of magnitude: the up to date experimental uncertainty of JVS systems is limited to a few parts in $10^{10}$ [4]. Therefore, a way to realize a QMT that avoided the practical difficulties arising from the limited SET current levels was needed.

In 1992, a pioneering work from NIST (USA) formulated for the first time a practically realizable QMT experiment [28] (Fig. 2 a). The key idea is accumulation of the charge delivered by a SET pump on a cryogenic capacitor, mounted in a dilution refrigerator system in close proximity with the SET device. For a suitably small capacitance value $C$ - typically in the pF range - integration of the SET current over a period of a few tens of seconds creates a reasonably high voltage $U$ - typically in the range of few volts - across the capacitor electrodes. Furthermore, an SET electrometer was also introduced in the experimental scheme control the charging process of the capacitor.

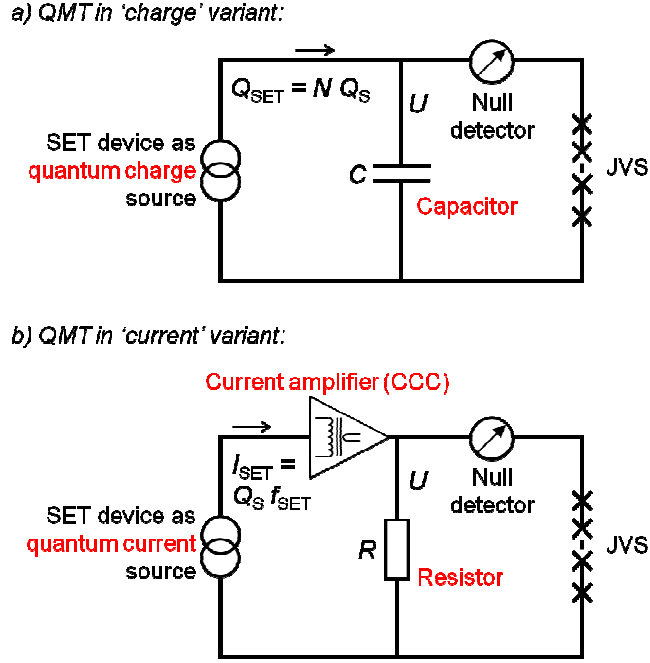

*Fig. 2: Basic principles of the two relevant variants of QMT experiments in simplified, schematic representations: Panel a) shows the 'charge' (ECCS, or indirect) QMT variant, and panel b) the 'current' (Ohm's law, or direct) variant.*

It is straightforward to understand that this experiment is an equivalent representation of the QMT according to equ 1. For this, we first consider that the capacitance $C$ of the capacitor can be traced to $R_K$ via the QHR in the $i$th quantized resistance plateau by using a suitable quadrature impedance bridge working at a frequency $\omega$ according to

$$C = 1/(\omega R_{QHR}) = i/(\omega R_K) \qquad (4)$$

With $Q_{SET} = N\, Q_S$ being the total charge of $N$ electrons moved between the capacitor electrodes by the SET device, it follows from $Q_{SET} = C\, U$ that

$$N\, Q_S = (U\, i)/(\omega R_K) \qquad . \qquad (6)$$

Finally, by measuring $U$ using a JVS system according to equ i), we obtain $N\, Q_S = (i\, n f_J/K_J)/(\omega R_K)$, or

$$K_J\, R_K\, Q_S = (i\, n\, f_J)/(N\, \omega) \qquad (7)$$

Hence, this QMT variant relates the product of the phenomenological constants (left side of equ. 7) to a product of integer quantum numbers and a ratio of two frequencies (right side of equ. 7), similar to equ. 1. Consequently, all implications of equ. 2 and equ. 3 also hold here.

The QMT experiment in the 'charge' version was pursued by the NIST group and called an '*Electron Counting Capacitance Standard*' (ECCS) [24, 29, 30, 31]. This name indicates that in the beginning this experiment was meant to result in a new, quantum-based capacitance standard. Only about 8 years later were its implications interpreted more in terms of a QMT experiment. A few years later, similar capacitance-based QMT implementations were started at several European NMIs - NMI/VSL (NL), NPL (UK), OFMET/METAS (CH), and PTB (D) - and pursued in the frames of three joint European metrology research projects [32, 33, 34].

Another practical way to cope with the small SET currents in a QMT experiment was presented in 2000 by the fundamental electrical metrology group of BNM/LNE (FR) [11]. This proposed QMT implementation, schematically sketched in Fig. 2 b), is based on amplifying the SET current by at least a factor of 10 000 by using a cryogenic current comparator (CCC) coupled to a dc SQUID magnetic flux detector in a dilution refrigerator environment. The amplified current is fed through a standard resistor, traceable to a QHR, that acts as a current-voltage converter. The resulting voltage is then directly measured by the use of a JVS system. The set-up of this experiment was started at BNM/LNE and also pursued within the already mentioned joint EU projects [32, 33, 34].

Recent progress in SET current source devices [35, 36] has also motivated the development of such versions of 'direct' QMT experiments in which the amplification of the SET current by a high-gain CCC is not needed [37].

The relation between the two variants of the QMT, represented by equations 1) and 7) and shown in Fig. 2, is schematically shown in Fig.3.

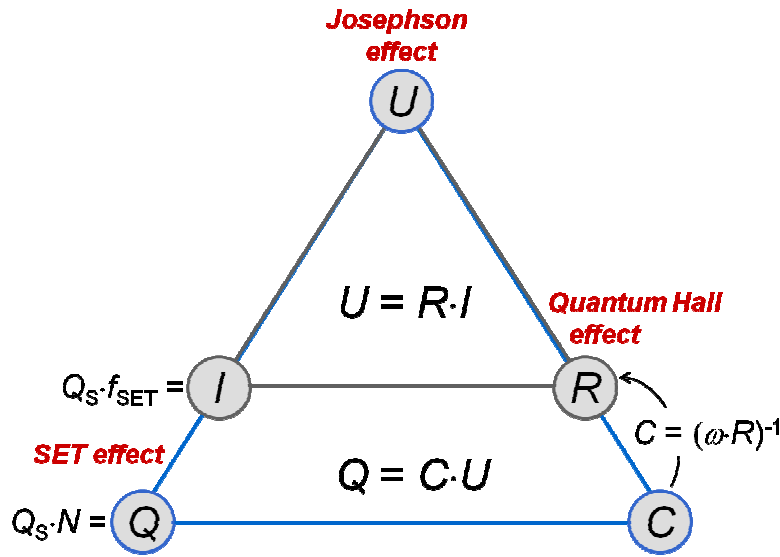

*Fig. 3: Extension of the original version of the QMT (the 'current' variant, cf. Fig. 1), based on the Ohm's relation U = R I, to the 'charge' variant, based on the capacitance relation Q = C U. The link to the QHR is given by the impedance of the capacitance C through a quadrature bridge.*

In summary, two distinct variants are used for experimental implementations of the QMT. The first one is the ECCS (Fig. 2 a), which uses the SET device as a quantum charge source and relies on a cryogenic capacitor. The second variant exploits Ohm's law (Fig. 2 b). It requires a suitable resistor and typically a CCC-based current amplifier with sufficiently high gain, and is also called the 'current' or 'direct' variant of the QMT. Both versions of the QMT include a null-detector instrument, whose performance typically limits the achievable accuracy of these experiments. In the 'charge' variant, an SET electrometer is used as a null-detector in the voltage feedback of the setup (see Fig. 2a and Fig. 7). The noise and drift of the electrometer, caused by the dynamics of background charges in the vicinity of the SET device, set the accuracy limit of the experiment, while in the 'current' variant the noise level of the SQUID detector involved in the CCC amplifier typically sets the limits.

QMT experiments based on the ECCS have already produced results at NIST [24, 25, and references therein] and at PTB [38, 39, and references therein]. QMT implementations of 'current' variant are being developed by two European NMIs, which are LNE (FR) [11, 40, 41 and references therein] and MIKES (FI) [37, 42 and references therein]. The status and results of these experiments are discussed in more detail later in this paper.

Several new approaches to implement direct (or 'current') QMT variants are presently under development at some NMIs, among them NPL (UK) and PTB (D) which involve measurements of the SET current by advanced current-voltage conversion methods, for instance by using high-ohmic resistors, traceable to the QHR [43]. However to date this is still work in progress, and the experiments have not yet delivered significant results with respect to the QMT. Another, even more ambitious approach for a future QMT realization was recently proposed in [44]. This idea is based on the monolithic integration of GaAs-based QHR and single-electron pump devices on a single chip.

## 5. Note on the SET "leg" in QMT experiments

Any QMT experiment requires that all relevant experimental parameters have to be well controlled to assure proper operation conditions for the electrical quantum effects involved. For the QHR, used to provide the link of the QMT resistance or capacitance 'leg' to $R_K$, as well as for the JVS, linking the voltage leg to $K_J$, this is feasible by applying well-established methods that are common in modern metrology laboratories. In addition to the bias current applied to the quantum electrical devices and the system temperature, other relevant parameters are the magnetic inductance applied to the QHR device, and the microwave frequency $f_J$ irradiating the Josephson contact in the JVS. The exactness of the QHR and JVS benefits from the fact that the relevant experimental parameters are relatively easy to control in practical applications, as well as from the rather 'robust' nature of the underlying macroscopic quantum effects.

It is important to note that this is more difficult for the SET leg in QMT experiments where SET devices are used as current or charge standards. The preparation of their proper operation conditions is typically less straight-forward and more complex. It generally requires:

- sub-Kelvin cryogenic environment by the use of dilution or He3 refrigerator systems,
- thorough shielding of the SET devices from thermal background radiation, and

- extremely careful low-pass filtering of the experimental wiring to avoid electromagnetic rf interference effects; for typical SET current sources an attenuation of about 100 dB for frequencies of 1 GHz and above is required [45, 46, and references therein].

Special challenges arise not only because the nanometer-scale SET devices are electrically very fragile circuits that can easily be destroyed by handling during an experiment; but also because they are more susceptible to intrinsic error effects, due to the sensitivity of the underlying microscopic Coulomb blockade effects [30, 31, 38, 39, 40, 41, 47]. In a 'current' QMT experiment, for example, the SET-generated current is described by the relation

$$I_{SET} = \langle n_{SET} \rangle \cdot Q_S \cdot f_{SET}, \tag{8}$$

where $f_{SET}$ is the driving frequency applied to the SET device and $\langle n_{SET} \rangle$ is the average number (over many clock cycles) of charge quanta transferred per clock cycle with repetition frequency $f_{SET}$. In the ideal case, $\langle n_{SET} \rangle = 1$ for normal metallic and semiconductor SET devices, or $\langle n_{SET} \rangle = 2$ for superconducting devices which pump Cooper pairs.

In a real experiment, however, $\langle n_{SET} \rangle$ typically deviates from the ideal value due to error effects. Such errors typically occur randomly in time during operation, and can, for instance, be caused by co-tunneling, by other parasitic tunneling events or by 'missed cycle' events, and can also be triggered by rf background interference or by thermal activation [9, 27, 45, 46 and references therein]. Depending on the quality of the experimental setup and the setting of the SET device operating parameters ('tunig'), this can lead to deviations from the ideally quantized behavior amounting to parts in $10^6$ or more [40, 41, and references therein]. Proper tuning of SET device requires in particular the adjustment of their working point via external control parameters, typically dc and ac voltage levels on gate electrodes of the SET device. The same applies when SET devices are used as quantum charge sources ($Q_{SET} = N\, Q_S$) in ECCS experiments [30, 31, 45, 47, and references therein].

In light of what has been discussed so far, it follows that an indispensable prerequisite for the metrological application of SET devices is the quantitative verification of their single-electron transfer accuracy. Consequently for any QMT experiment, by definition aiming at a consistency check of the phenomenological constants, this verification must be done by means *independent* from SET current or charge measurements which are parts of the QMT. In this respect, for instance, an interesting finding from the LNE experiments was that deviations of $I_{SET}$ from the expected quantized value $e \cdot f_{SET}$ were observed although the measured current plateaux (i.e. $I_{SET}$ plotted versus pump bias voltage) showed reasonable flatness [41, 8]. This strongly indicates that current plateaux flatness has to be considered a *necessary, but not sufficient* indication for the proper operation of SET current source devices as quantum current standards.

In the ECCS experiment developed at NIST [24, 25, 28, 30] as well as in the experiment at PTB [38, 39, 48], the quantitative determination of SET error effects was carried out by performing a preliminary 'shuttle pumping' experiment. The SET device is connected to an on-chip metallic island provided with a small stray capacitance $C_{stray}$ (Fig. 4). This node is via a coupling capacitance $C_{cp}$ electrostatically connected to the input of an SET electrometer, which provides sub-$e$ charge resolution, as the ratio $C_{cp}/C_{stray}$ is made sufficiently large by a suitable device design. The SET pump is operated so that it repeatedly pumps one electron in and out from the island, while the electrometer is used to monitor the charge state of the island.

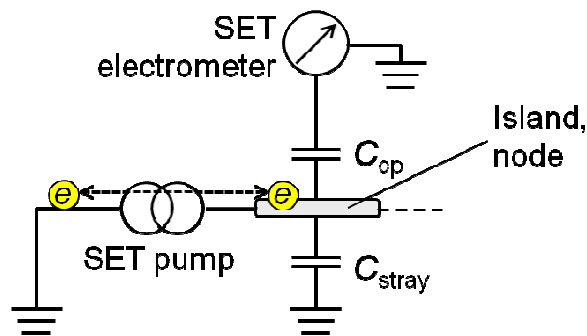

*Fig. 4: Schematic operational principle to detect transfer errors in an SET pump device by the 'shuttle pumping' method. Single electrons are shuttled in and out from the island at the frequency $f_{SET}$. If the effective charge divider ratio $C_{cp}/C_{stray}$ is sufficiently large (typically around 1/20), single electron charges on the island can be resolved by the SET electrometer, and pump error events can be detected.*

The charge transfer accuracy of the SET pump is determined by measuring the average rate of the error events detected by the electrometer and relating this number to the pumping frequency $f_{SET}$. The best results for SET pumps based on metallic tunnel junctions, obtained for frequencies $f_{SET}$ of few MHz, correspond to relative single-electron transfer errors of about few parts in $10^8$ or better [24, 25, 30, 48].

Although this procedure, the first and still only SET error detection method applied successfully in QMT experiments, is suitable to quantify SET error contributions for uncertainty assessments [24, 25, 30, 48], it has conceptual flaws and limitations. The main one is given by the necessary assumption that the error rates during the shuttling phase, determined by bidirectional pumping of single electrons, are equal to the ones in the unidirectional pumping process phase of the experiment, when the SET current (or charge) is sourced to the resistor in a 'current' QMT (or to a capacitor in a 'charge' QMT). More advanced variants for SET error detection and accounting is currently pursued at PTB [49, 50]. Here, the errors occurring in a serial array of (two or more) SET pump devices are detected on small charge nodes between each two pump devices by using SET electrometers as single-electron charge detectors. A logic circuit, processing the output signals of individual electrometers, allows then to identify the error-producing device. Once errors are identified and quantified, they can be incorporated as known correction terms for the determination of the current or charge sourced by the device.

Reliable error detection requires SET electrometers with sufficiently large bandwidth. For very well-performing pump devices with error rates of the order of about 100 s$^{-1}$ (corresponding, for example, to an accuracy of 1 part in $10^6$ at a pumping frequency $f_{SET} = 100$ MHz), conventional dc-SET electrometers are still adequate. For less accurate devices with higher error rates, or for higher pumping frequency $f_{SET}$, the detector must have a correspondingly larger bandwidth. This is achievable by the implementation of an rf-SET circuit operating with a typical carrier frequency of about 500 MHz and with a bandwidth around 1 MHz [51].

Several European institutes, among them the NMIs MIKES (FI), NPL (UK) and PTB (D), will pursue the development of advanced SET error accounting schemes in a new joint European research project throughout the next years [52].

## 6. Results and progress of QMT experiments worldwide

In the following, the principles, preliminary results and ongoing progress of existing QMT experiments is reviewed, each including an assessment of the estimated ultimate accuracy limit of the experimental variant.

### 6.1. Direct' QMT experiments

#### The 'direct' QMT experiment at LNE

The QMT experiment at LNE uses a 3-junction $R$-pump, developed and fabricated at PTB, for the SET current generation, and a specially developed CCC for amplifying this current. A simplified scheme of the experimental setup is shown in Fig. 5. The experiment is described in [11, 40, 41, 8, 53] and in references therein. Further details and recent results are published in a dedicated article by the LNE group in this journal issue [54].

The $R$-pump is an improved concept of the conventional SET pump based on Al-AlO$_x$-Al tunnel junctions [55]. This pump is equipped with on-chip chromium micro-strip resistors in series with the junctions, each resistor having a resistance exceeding $R_K$. The resulting modification of the effective electromagnetic environment of the junctions has been shown to suppress unwanted co-tunnelling events, which are presumed to compromise the accuracy of $I_{SET}$. At LNE the 3-junction $R$-pump was operated up at a maximum frequency $f_{SET} = 100$ MHz, corresponding to $I_{SET} = 16$ pA. The current amplifier of LNE was composed of a CCC with a high winding ratio $G = 20\,000:1$ together with a dc-SQUID, capable of amplifying $I_{SET}$ to about 0.3 µA. A secondary current source is servo-controlled by the SQUID which works as a null detector for the magnetic flux $\Phi$ in the CCC. The polarity of the SET current to be amplified is periodically reversed in order to reduce contributions from $1/f$ flicker noise. The voltage $U$ across the room-temperature standard resistor ($R = 10$ kΩ) is simultaneously measured by a programmable JVS system in combination with a precision voltmeter. The irradiation frequency $f_J$ of the JVS and also the pumping frequency $f_{SET}$ are referred to a 10 MHz frequency standard.

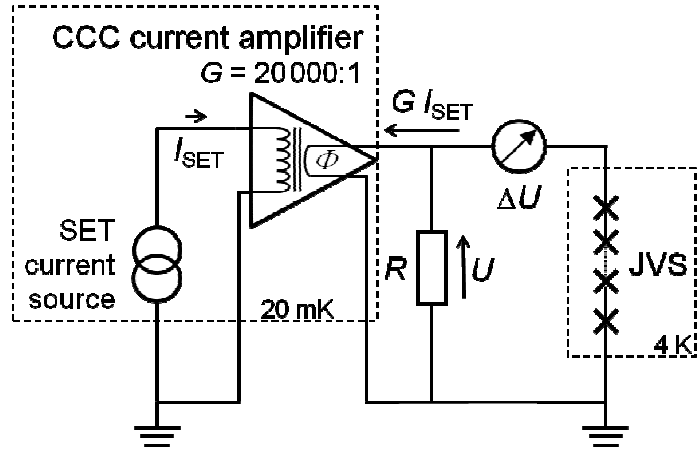

*Fig.5: Principle of the QMT experiment at LNE, involving a CCC-based current amplifier and an R-pump as SET quantum current source. Two detectors are used: a SQUID (not shown) which detects and nulls the magnetic flux $\Phi$ induced in the CCC, and a voltmeter to measure the deviation $\Delta U$ from the quantized voltage given by the JVS system.*

The preliminary results of the LNE experiment presented in [41, 8, 53] showed a relative deviation $\Delta_{QMT}$ from the expected QMT relation (equ. 3) of few parts in $10^4$, with a relative uncertainty of $u_{QMT}$ of few parts in $10^6$. Considering the fact that the experiment suffered from irreproducibility problems observed in a series of measurements [41, 8, 53], and given the present knowledge on the maximum $\Delta_{QMT}$ value to be expected, which is less than one part in $10^6$ [7], those preliminary results hinted to problems of the experiment. However, very recently improvements of the setup remedied the lack of reproducibility, and the best result achieved in the LNE experiment to date is $Q_S/e - 1 = (-5 \pm 13) \cdot 10^{-6}$ [54].

The measurement uncertainty for the LNE QMT experiment is in principle limited by statistical (type A) uncertainty contributions dominated by the noise of the SQUID null detector. These uncertainty contributions are inversely proportional to the current and inversely proportional to the square root of the measurement time. The largest uncertainties related to systematic effects (type B components) are estimated to be on the order of one part in $10^8$ or less, and depend weakly on the current level [34]. They arise from the CCC ($u_{CCC} \cong 10^{-8}$ including capacitive leakage, finite open loop gain and winding ratio error), the calibration of the 10 k$\Omega$ standard resistor against a QHR (typically $u_{QHR} < 10^{-8}$), and the systematic uncertainties related to the JVS system ($u_{JVS} < 10^{-8}$, mainly due to residual thermal voltages, resistive leakage, and detector and frequency errors). However, the QMT experiment at LNE lacks of the means for an independent determination of the SET transfer errors, e.g. by shuttle pumping measurements, since it does not include a single-electron charge detector. Thus, uncertainty contributions related to the SET pumping errors cannot be quantified.

The principal ultimate accuracy limit of the experiment, assessed in frame of the REUNIAM project [34], is crucially dependent on the performance of the CCC including SQUID detector. For $I_{SET} = 1$ pA and a CCC input current resolution of 1 fA/√Hz it was estimated that a standard uncertainty of about 4 parts in $10^6$ should be realistically achievable during a measurement time of 10 hours. Considering a relative standard uncertainty of one part in $10^8$ as the ultimate, ambitious target for QMT experiments, it was further concluded that the LNE experiment could be performed with such uncertainty if the following conditions are fulfilled:

- availability of a CCC with a current resolution of 1 fA/Hz$^{1/2}$ or less in the white noise regime,
- immunity of the electrical wiring between the CCC and the SET current source against microphonic and interference pick-up effects, and
- availability of an SET current source generating $I_{SET} \geq 100$ pA with highly stable performance.

### The 'direct' QMT experiment at MIKES

The QMT experiment currently under development at MIKES will involve a hybrid turnstile device as SET current source, a cryogenic resistor, and a cryogenic current null detector.

Hybrid turnstiles are a relatively new kind of SET quantum current source devices [36]. They comprise two metallic nano-scale superconductor–insulator–normal (in this sense 'hybrid') tunnel junctions in series. The interplay of the Coulomb blockade and the superconducting energy gap enables the clocked transfer of single electrons by using only one driving gate signal. The devices are categorized as 'turnstiles' since they must be operated with a finite bias voltage applied to their source-drain terminals, in contrast to pumps which are able to clock-transfer electrons without such bias [9].

In the QMT experiment at MIKES, the current $I_{SET}$ delivered by the turnstile device will be directly opposed to a current $I_R$ which is generated by applying a Josephson voltage to a cryogenic resistor with a resistance of 1 MΩ [37, 42]. The small unbalanced current difference $\Delta I = I_{SET} - I_R$ is detected by a cryogenic null detector, presently realized by a dc current transformer with moderate gain in combination with SQUID as current null detector. The principle of the experiment is sketched in Fig. 6.

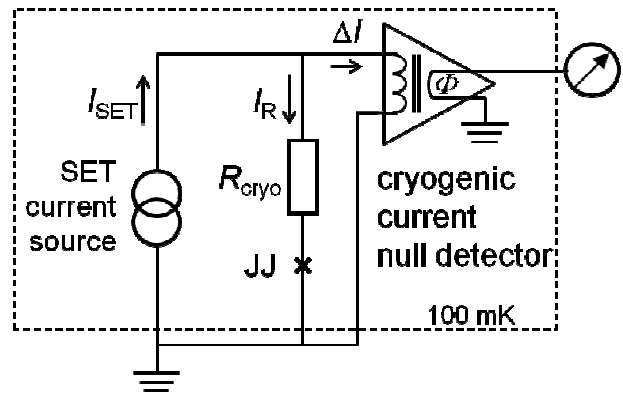

*Fig. 6: Principle of the QMT experiment under construction at MIKES. The current $I_{SET}$ delivered by the SET device is nearly balanced by an opposite current $I_R$ from the voltage of a Josephson junction biasing a cryogenic resistor ($R_{cryo}$ = 1 MΩ). The residual current difference $\Delta I$ is measured by a cryogenic null detector including a dc current transformer and a SQUID (not shown).*

In the first experiments it is planned to operate the cryogenic resistor at a temperature of 0.7 K, which creates a Nyquist current noise of about 6 fA/√Hz [34]. The dominant type-A uncertainty contribution in this experiment will, however, be given by the noise level of the SQUID null detector, generating a noise equivalent to 20 fA/√Hz or higher. In a later development stage, this will be improved by using a null detector specially designed for this purpose. Assuming that the SET current device is generating a current of 100 pA at sufficient accuracy, the noise figures of the setup would limit the total relative uncertainty to about 8 parts in $10^7$, requiring an averaging time of about 10 h. In the case if the hybrid turnstile current source would be able to produce $I_{SET}$ = 100 pA at sufficient accuracy, the uncertainty could be reduced to about 8 parts in $10^7$. These preliminary estimates have neglected possible flicker noise and drift effects which may appear when measurements are averaged over a very long time.

A preliminary assessment of the possible ultimate accuracy limit of this experiment shows that - besides the noise of the current null detector - the current dependence and possible flicker noise of the thin-film cryogenic resistor are the dominant type-A contributions [34]. The calibration of this resistor against a QHR at a current of about 1 µA is possible with a relative uncertainty $< 10^{-7}$ if a CCC bridge is used, but difficulties may arise since the maximum current of the SET device is limited to about 100 pA. This mismatch in current together with the current coefficient of the cryo-resistor may cause relative uncertainties of the order of few parts in $10^6$ [56].

A significant improvement of the uncertainty of this QMT experiment below 1 part in $10^6$ would require a better understanding of these current dependence effects and the availability of a null detector with lower noise floor. In addition it would need a drastic increase of the output current of the quantum current source by about a factor of 10 to reach the 1 nA level, which seems not possible at present but may be feasible in future, e.g. by a parallel combination of SET current source devices.

### 6.2. 'Indirect' QMT experiments

**The ECCS experiment at NIST**

After the invention of the principle for the Electron Counting Capacitance Standard experiment in 1992 [28], the Martinis group at NIST continuously developed a corresponding experiment. In the beginning, their work was focussed on the development of a suitable SET pump device, starting with metallic single-electron pump containing five junctions in series [29]. Since its pumping accuracy was found to be insufficient for the metrological purpose, in the following years pumps with an increased number of junctions were developed and investigated. In 1996, the first 7-junction pump with sufficiently high pumping accuracy, i.e. with a relative uncertainty of only about 1.5 parts in $10^8$, was presented [30, 31]. Such pump was used in the 1999 experiment which demonstrated the first proof-of-principle of the ECCS [24].

Besides the 7-junction SET pump combined with an SET electrometer on-chip, the NIST experiment comprised a vacuum-gap cryogenic capacitor ($C_{\text{cryo}} \cong 2$ pF, in the following for simplicity called 'capacitor') with parallel-plate arrangement of the electrodes. Furthermore, two specially designed mechanical needle switches were used to provide switchable electrical contacts between the SET chip and the capacitor, or, respectively, between the capacitor and a capacitance bridge for measuring $C_{cryo}$. The experimental setup is schematically shown in Fig. 7 and in detail described in [24, 25, and references therein].

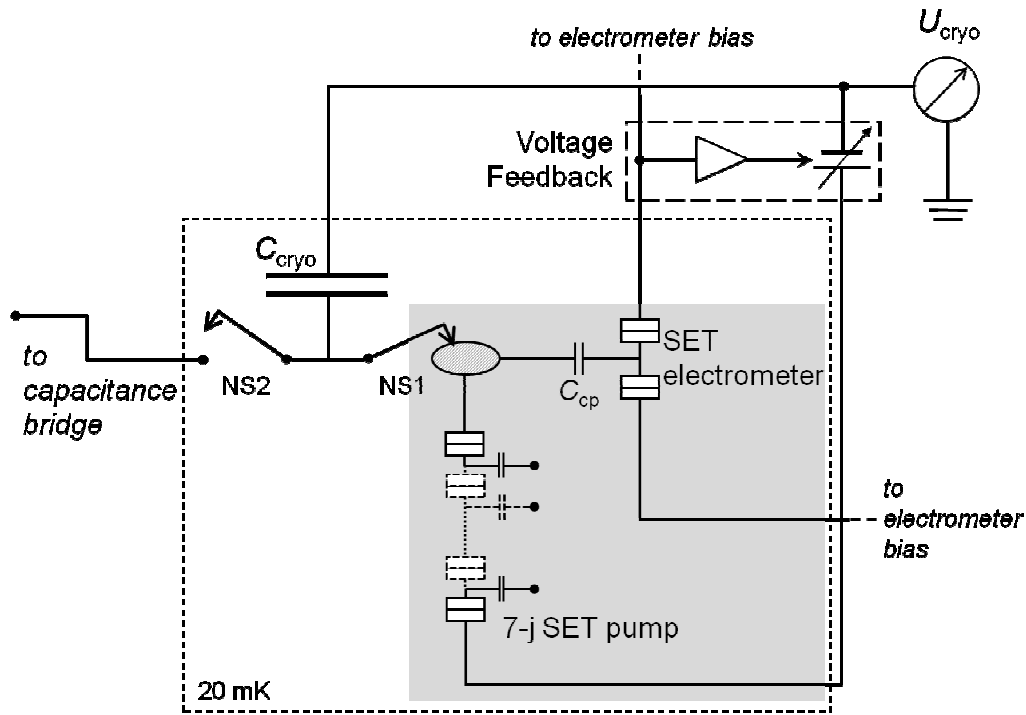

*Fig. 7: Scheme of the ECCS experiment at NIST. The chip with the SET circuit (7-junction SET pump and SET electrometer) is shown shaded. In the capacitor charging phase shown here, the needle switch NS1 is closed to connect the SET pump to the cryogenic capacitor while the SET electrometer acts as voltage null detector, controlling the servo voltage $U_{cryo}$ driven by the feedback circuit. In the next phase, the needle switch NS1 is opened and NS2 is closed to connect the cryogenic capacitor to a capacitance bridge for measuring $C_{cryo}$. The bias circuit for the electrometer source/drain terminals (dotted line ends in the figure) is not shown for clarity.*

After tuning the SET pump for its optimum working point, i.e. adjusting the dc voltages on the pump gate lines to the six pump islands for minimizing pumping errors during shuttle-pumping, the ECCS experiment is performed according to the following procedure.

With the needle switch NS1 closed, the SET pump transfers charge quanta onto one side (the 'low potential' electrode) of the capacitor. In order to maintain proper working conditions for the pump during this phase, the voltage across the pump must be kept near zero. This also ensures that all transferred charge is collected on the capacitor electrodes and not on the stray capacitances between the SET chip and ground (not shown in Fig. 7). This is done by using the electrometer as a null detector for driving a feedback circuit that applies a compensating voltage $U_{\text{cryo}}$ to the 'high potential' electrode of the capacitor. The feedback voltage $U_{\text{cryo}}$ is constantly measured by using a high-resolution voltmeter, which is calibrated with a JVS. Typically, the capacitor is charged up to about 10 V while monitoring $U_{\text{cryo}}$ during several successive charging-discharging ramp cycles. Details on the experimental ECCS procedure as well as on the data analysis are given in [24, 25].

The operation of the initial prototype ECCS experiment, reported in 1999 [24], showed a reproducibility of order of $10^{-7}$ (relative scatter of the result data), but first lacked a full uncertainty analysis. The completion of the uncertainty budget required quantifying several Type B uncertainties, particularly the frequency dependence of the cryogenic capacitor, which was accomplished in 2006 [57]. The full uncertainty budget for this first ECCS experiment (nicknamed ECCS-1) was published in 2007 [25], and the result was

$$(\Delta_{\text{QMT}} \pm u_{\text{QMT}})_{\text{ECCS-1}} = (-0.10 \pm 0.92) \cdot 10^{-6}. \tag{9}$$

Thus, the ECCS-1 experiment 'closed' the QMT ($\Delta_{\text{QMT}} < u_{\text{QMT}}$) with a relative uncertainty of about $0.9 \cdot 10^{-6}$, which was the first result of any QMT experiment ever realized, and is still the best result for any QMT experiment to date.

The further analysis in [25] showed that the achievable uncertainty of the ECCS-1 was determined by the calibration uncertainty of the commercial capacitance bridge used, which was traced to the calculable capacitor of NIST. An improved setup for a second generation of the ECCS experiment was announced by the NIST group which should be able to overcome this limitation as well as others, and finally allow the realization of an ECCS that could achieve a total uncertainty of about $3 \cdot 10^{-7}$. In the following years, the development of such improved setup was pursued at NIST, and extensive practical knowledge on the operation of the ECCS was gathered [45]. However, due to technical problems with the fabrication of suitably accurate SET pumps, the successful implementation and execution of an improved ECCS experiment was not completed, and finally NIST stopped work on this experiment in 2008.

The combination of the ECCS-1 experiment result (equ. 9) with those of a Watt balance experiment was discussed in [26]. This combination forms a QMT that yields a value for $Q_S$ in terms of the SI coulomb, independent of the Josephson and quantum Hall effects. The result was

$$Q_S/e - 1 = (- 0.09 \pm 0.92) \cdot 10^{-6}, \qquad (10)$$

with an uncertainty identical to that of the ECCS-1 experiment.

In summary, the best knowledge to-date about the QMT is represented by the ECCS-1 experiment from NIST, implying that the validity of the relation $R_K \cdot K_J \cdot Q_S = 2$ is experimentally proven with an uncertainty of about 9 parts in $10^7$. Furthermore, it allowed to derive the value of the correction parameter $\varepsilon_S$ for the SET charge quantum, which was consistent with zero at the same uncertainty level [26].

**The ECCS experiment at PTB**

The ECCS experiment pursued at PTB is similar to the original NIST setup (see Fig. 7), however it differs in significant points (see [38, 39, 48] and references therein):

i) The SET quantum charge device is of $R$-pump type mentioned above [55]. However, instead of a 3-junction device as used in the direct QMT experiment at LNE, the ECCS at PTB uses a 5-junction $R$-pump which has shown relative single-electron transfer errors corresponding down to only few parts in $10^8$ in shuttle-pumping characterization measurements [39, 48]. Given the fact that this pump only needs four gate electrodes to be tuned for adjusting the working point (corresponding to the four pump islands each between two of the 5 junctions in series), the practical benefit of this pump is its easiness in use (compared to a 7-junction pump as used by NIST [24]) without sacrificing too much performance in pumping accuracy.

ii) The cryogenic vacuum gap capacitor used in the PTB experiment (in the following called 'capacitor' for simplicity) has a coaxial electrode arrangement with a capacitance $C_{cryo} = 1$ pF [58]. Trimming of the capacitor electrodes allowed to tune $C_{cryo}$ to the decadic value of 1 pF within $10^{-5}$ (relative deviation). The robustness of the coaxial construction resulted in a reproducibility of $C_{cryo}$ of about $10^{-5}$ (relative scatter) between thermal cycles, which allows high-precision capacitance measurements by the use of special bridge techniques [59, 38, 39]. Furthermore, the larger distance between the capacitor electrodes (5 mm for the PTB design vs. 50 µm for the NIST design) makes the frequency dependence of $C_{cryo}$ smaller than two parts in $10^8$ [38].

iii) A high-precision capacitance bridge technique, developed and available at PTB [59], allows $C_{cryo}$ to be measured in terms of $R_K$ with unexcelled accuracy of few parts in $10^8$. Thus, the dominant uncertainty contribution in the ECCS-1 uncertainty budget [25] will be negligible in the final PTB experiment.

After a significant improvement of the SET chip design, first preliminary results of the ECCS experiment at PTB were published in 2012 [39]. A full uncertainty budget was not available because several Type-B uncertainties have not been quantified yet, but the conservative estimation of their contributions allowed the quantification of a preliminary result (nicknamed ECCS-2)

$$Q_S/e - 1 = (-0.31 \pm 1.66) \cdot 10^{-6}. \tag{11}$$

Like the ECCS-1 from NIST, this result is also consistent with zero and, thus, 'closing' the QMT, however with a still slightly higher relative uncertainty of about 1.7 parts in $10^6$.

The conditions for this ECCS experiment are not completely optimized to date, and further improvements of the PTB experiment are currently pursued. It is expected that the total uncertainty eventually can be reduced to 3 parts in $10^7$ [39, 48]. Since the publication of [39], significant progress in the improvement of the pumping accuracy and in the JVS-based voltage measurement of $U_{cryo}$ already has been achieved [48]. Once all further improvements are implemented, the ECCS experiment at PTB is expected to produce results with an uncertainty level of down to three parts in $10^7$.

A principle accuracy limitation of the ECCS experiment remains in the type-B uncertainty component given by the ac-dc difference $C_{cryo}(f)$ of the cryogenic capacitor. Such dependence has to be considered in the frequency range from about 10 mHz, which is the effective frequency of the capacitor charging cycles in the ECCS, up to about 1 kHz, which is the typical operating frequency of the capacitance bridge. The crucial point here is that to date no experimental measurement techniques exist which allow the determination of this frequency dependence with the necessary accuracy, i.e. with a relative uncertainty of better than $10^7$. All ECCS experiments performed yet thus rely on estimates for the frequency dependence of the capacitors involved [25, 38, 39] which are based on reasonable model assumptions [57]. A corresponding conservative estimate for the PTB capacitor implies a very small frequency dependence of about 2 parts in $10^8$ or less [38], but the experimental verification still remains a task of paramount difficulty.

## 7. Discussion

Fig. 8) summarizes the results from the 'indirect' QMT experiments at NIST (ECCS-1) and PTB (ECCS-2) in terms of the measured values for $Q_S/e - 1$, together with two corresponding values derived from CODATA values through different routes.

The corresponding value from the ECCS-1 experiment [24, 25] stems from [26] where the ECCS-1 result was combined with results from Watt balance and calculable capacitor experiments. The corresponding figure derived from the preliminary result of the ECCS-2 experiment stems from [39]. The best result of the LNE 'direct' QMT experiment ($Q_S/e - 1 = -5 \pm 13 \cdot 10^{-6}$ [54]) is not shown here because it is not within the scale of the graph.

The data point with the value $2/(K_J \cdot R_K \cdot e) - 1 = -9,5 \cdot 10^{-10}$ shown in the left panel of Fig. 8) was derived by using actual CODATA values for $K_J$, $R_K$ and $e$ [21], however considering the corresponding uncertainty via two different routes, shown by the two different error bars:

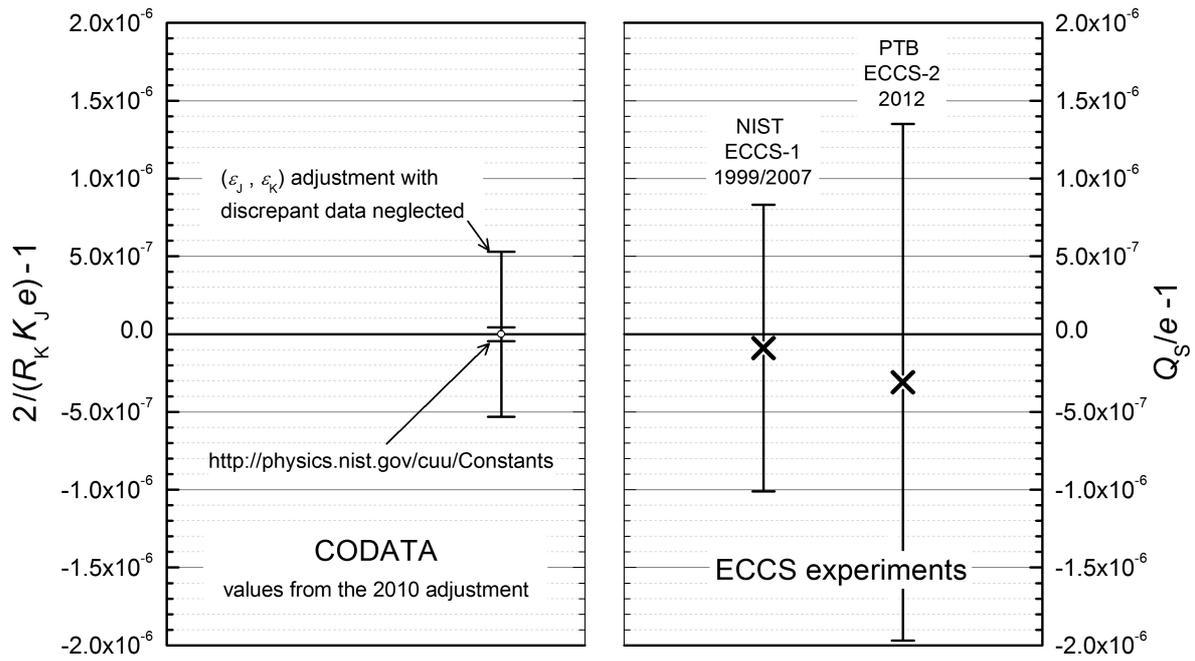

*Fig.8: Right panel: Results of the QMT experiments at NIST (ECCS-1) and PTB (ECCS-2) in terms of the measured values for $Q_S/e - 1$. Left panel: CODATA value (data from the 2010 adjustment) for the corresponding value $2/(K_J \cdot R_K \cdot e) - 1$ with error bars from two different routes for the uncertainty analysis (see text).*

- the smaller error bar corresponds to the total standard uncertainty $u_{\rm rel}(2/(K_J \cdot R_K \cdot e)) \cong 4.4 \cdot 10^{-8}$ when the actual CODATA standard uncertainties for $K_J$, $R_K$ and $e$ are used [21], which are $u_{\rm rel}(K_J) = 2.2 \cdot 10^{-8}$, $u_{\rm rel}(R_K) = 3.2 \cdot 10^{-10}$ and $u_{\rm rel}(e) = 2.2 \cdot 10^{-8}$. The effects of possible correlations between these uncertainty values, inherent in the CODATA analysis, were neglected here.

- the larger error bar corresponds to the total uncertainty $u_{\rm rel}(2/(K_J \cdot R_K \cdot e)) \cong 5.3 \cdot 10^{-7}$. This figure results when the uncertainties for $K_J$ and $R_K$ from the 2010 CODATA adjustment under 'relaxed conditions' and with discrepant input data neglected are used ($u_{\rm rel}(K_J) = 49 \cdot 10^{-8}$ and $u_{\rm rel}(R_K) = 1.8 \cdot 10^{-8}$) [6].

Fig. 8) shows that the impact threshold regarding the uncertainty level of QMT experiments is dependent on the interpretation of the latest CODATA adjustment results. An assessment based on the rigorous CODATA analysis, i. e. on the assumption that the fundamental relations ib) – iiib) are exact, indeed implies that an uncertainty level of the order of one part in $10^8$ is necessary to provide significant input to future adjustment calculations of the constants. However, the more conservative approach, i. e. considering results from an adjustment under 'relaxed conditions' according to ic) – iiic), implies that impact is possible at an uncertainty level of 5 parts in $10^7$ or less.

## 8. Conclusion and outlook

More than two decades of experience with different setups of QMT experiment at several NMIs worldwide has shown that their setup requires overcoming manifold difficulties and practical challenges, and therefore long-term efforts. This is not only because of the special challenges with the operation of SET devices at a metrological accuracy level; rather it is also because of the very nature of the QMT in which all three quantum electrical standards must be combined properly, and operated linked together. As discussed in this article, the total uncertainty of the QMT experiments, pursued currently and in the past, may be reduced down to a few parts in $10^7$ as a realistic target with the present methods, provided that all feasible improvements in the set-ups are implemented successfully.

For the 'direct' or 'current' type QMT experiments, involving a high-gain CCC or a cryogenic null detector, respectively, the most important condition is the availability of robust and highly stable SET current sources. These devices must be capable of delivering SET currents exceeding 100 pA significantly. Other obstacles remain to be overcome, particularly the reduction of the white noise floor of the complete system, corresponding to a current noise level of down to 1 fA/√Hz or less.

The 'indirect' or 'charge' type QMT experiment, aka ECCS, at PTB has the potential to reach a total uncertainty of 3 parts in $10^7$ after the completion of further improvements in reach and when all experimental components are operating properly [39, 48]. A result at this level would bear on possible corrections to both the SET charge quantum $Q_S$ and the Josephson constant $K_J$.

To date all relevant realizations of QMT experiments, reviewed in this paper, seem to cluster near an uncertainty level of about one part in $10^6$. However in the past years and ongoing, significant efforts and progress on the further improvements of all quantum electrical effects were made, in particular in the field of single-electron transport devices and their metrological application. This supports the expectation that QMT experiments pursued at several NMIs in the near future will be capable of reaching an uncertainty level of few parts in $10^7$, and so can produce relevant results for fundamental metrology. The ultimate target to close the QMT at an uncertainty level of about one part in $10^8$, however, remains a formidable experimental challenge.

## Acknowledgments

The authors gratefully acknowledge

M. W. Keller, A. Manninen, P. J. Mohr, D. Newell and F. Piquemal for their helpful comments on the manuscript, and

F. J. Ahlers, M. W. Keller and J. M. Martinis for valuable discussions and support on the ECCS experiment at PTB.